\documentclass[aps,prd,preprint,floatfix]{revtex4}
\usepackage{graphicx}
\newcommand{\be}{\begin{equation}}
\newcommand{\ee}{\end{equation}}
\newcommand{\bea}{\begin{eqnarray}}
\newcommand{\beas}{\begin{eqnarray*}}
\newcommand{\eea}{\end{eqnarray}}
\newcommand{\eeas}{\end{eqnarray*}} 
\newcommand{\ba}{\begin{array}}
\newcommand{\ea}{\end{array}}
\newcommand{\bi}{\begin{itemize}}
\newcommand{\ei}{\end{itemize}}
\newcommand{\ben}{\begin{enumerate}}
\newcommand{\een}{\end{enumerate}}

\newcommand{\gtsim}{\stackrel{\scriptscriptstyle>}{\scriptscriptstyle\sim}}
\begin{document}
%@twocolumnfalse\endcsname

\preprint{\vbox{
\rightline{BUHEP-02-40}
\rightline{FSU-HEP-021202} 
\rightline{IFUAP-HEP-03-02}
}}

{\tiny .} \vspace{1.5cm}

\title{Where is the Higgs boson?}

\author{A. Aranda}
\email[]{fefo@cgic.ucol.mx}
\altaffiliation{Present address: Facultad de Ciencias, Universidad de Colima, M\'exico.}
\affiliation{Department of Physics, Boston University, Boston, MA 02215 U.S.A}

\author{C. Bal\'azs}
\email[]{balazs@hep.fsu.edu}
\affiliation{Department of Physics, Florida State University, 
Tallahassee, FL 32306 U.S.A.}

\author{J.L. D\'\i az-Cruz}
\email[]{ldiaz@sirio.ifuap.buap.mx}
\affiliation{Instituto de Fisica, BUAP, Puebla, Pue. 72570, M\'exico}

\date{June 16, 2003}

\begin{abstract}
Electroweak precision measurements indicate that the standard model Higgs boson 
is light and that it could have already been discovered at LEP 2, or might be 
found at the Tevatron Run 2. In the context of a TeV$^{-1}$ size extra 
dimensional model, we argue that the Higgs boson production rates at LEP and the 
Tevatron are suppressed, while they might be enhanced at the LHC or at 
CLIC. This is due to the possible mixing between brane and bulk components of the 
Higgs boson, that is, the non-trivial brane-bulk `location' of the lightest 
Higgs.
To parametrize this mixing, we consider two Higgs doublets, one confined to the 
usual space dimensions and the other propagating in the bulk. Calculating the
production and decay rates for the lightest Higgs boson, we find that compared 
to the standard model (SM),
the cross section receives a suppression well below but an enhancement close to 
and above the compactification scale $M_c$. This impacts the discovery of the 
lightest (SM like) Higgs boson at colliders. 
To find a Higgs signal in this model at the Tevatron Run 2 or at the LC with 
$\sqrt{s}=1.5$ TeV, a higher luminosity would be required than in the SM case. 
Meanwhile, at the LHC or at CLIC with $\sqrt{s} \sim 3$--5 TeV one might find 
highly enhanced production rates. This will enable the latter experiments to 
distinguish between the extra dimensional and the SM for $M_c$ up to about 6
TeV.
\end{abstract}

\maketitle

%\pacs{PACS:}

%%%%%%%%%%%%%%%%%%%%%%%%%%%%%%%%%%%%%%%%%%%%%%%%%%%%%%%%%%%%%%%%%%%%%%

\section{Introduction} 
\label{sec:intro}

% Higgs
The Higgs boson is the missing link connecting the real world with the unified 
electroweak (EW) gauge group by spontaneously breaking the latter. Precision 
measurements of EW observables constrain the Higgs mass below about 200 GeV at 
95\% C.L.~\cite{Takeuchi:2003mk,Chanowitz:2003hx,Langacker:hi,Erler:2002ix,
Langacker:2002sy,Villa:2002zt,Chanowitz:2002cd} within the standard model (SM). 
Thus, it is expected that a Higgs particle will be 
discovered at the Run 2 of the Tevatron, provided sufficient 
luminosity~\cite{Carena:2000yx}.
But it is intriguing to notice that the EW observables strongly prefer a SM like 
Higgs with mass below 114.1 GeV~\cite{Langacker:2002sy,Chanowitz:2002cd}, which 
is the present lower limit from LEP 2. The data also indicate that the Higgs boson 
should have already been discovered~\cite{Langacker:2002sy}, and the fact that 
it was not found can be interpreted as new physics crucially affecting the 
Higgs sector~\cite{Chanowitz:2002cd}.
In this work we put forward a model in which the presently missing signal of the 
lightest Higgs boson is due to a suppression of the Higgs production cross 
section at LEP and the Tevatron. This suppression arises from the non-trivial 
`location' of the lightest Higgs boson in a five dimensional space. However, the 
same feature promises enhancement of the Higgs signal at the CERN 
Large Hadron Collider (LHC) and possibly at a multi TeV linear collider (CLIC).

% XD
%The five dimensional model that is used in this work arises as follows.
The idea that our universe could be confined to a higher dimensional defect 
has been revived both in field theory~\cite{Rubakov:bb} and string contexts
~\cite{Antoniadis:1990ew}. It has been laid on more solid ground in the context 
of non-perturbative string analyses~\cite{Polchinski:1995mt,Lykken:1996fj,
Horava:1996ma}, and applied as a possible solution to the gauge hierarchy 
problem~\cite{Antoniadis:1997zg,Arkani-Hamed:1998rs,Antoniadis:1998ig,
Arkani-Hamed:1998kx}. Such a solution relies on the existence of $n > 0$ 
additional compact space-like dimensions.
% hierarchy
In models based on this idea, the four dimensional Planck scale $M_{P\ell}$ 
becomes an effective quantity and it is related to the fundamental scale $M$ by 
the volume of the extra space $V_n$ via the relation $M_{P\ell}^2 = M^{n+2} 
V_n \label{mp}$ . If one requires $M = {\cal O}$(TeV) then for $n = 2$, the 
compactification radius $R \sim V_n^{1/n}$ is in the order of a millimeter, but 
for $n = 7$ it is less than a fermi, not far from the inverse of a TeV.
It is remarkable to notice that with ${\cal O}$(TeV$^{-1}$) size extra 
dimensions the hierarchy problem is indeed nullified, since the fundamental 
scales $M \sim 1/R$ are close to TeV. 

% TeV scale XD
The string arguments of Ref.s~\cite{Polchinski:1995mt,Horava:1996ma} 
also allow the standard gauge and Higgs sectors to penetrate the bulk. If the 
compactification scale $M_c = 1/R$ is higher than ${\cal O}$(TeV), 
phenomenology does not conflict with this scenario
either~\cite{Pomarol:1998sd,Antoniadis:1998sd,Delgado:1998qr,carone,Dienes:1998vh,Dienes:1999vg}.
This makes the inverse TeV size extra dimensional models attractive.
% or
Alternatively, the hierarchy problem can be solved with the use a non-% 
factorizable geometry, which has also been proposed in five dimensions. The 
introduction of an exponential `warp' factor reduces all mass parameters of 
the order of a fundamental $M_{P\ell}$ of a distant brane to TeV's on the brane 
where we live~\cite{Randall:1999ee}. In order to explain large hierarchies among 
energy scales one simply has to explain small distances along the extra 
dimension, thus this mechanism requires a Planck scale size extra 
dimension~\cite{Randall:1999ee,Arkani-Hamed:1999hk}.

These intriguing possibilities have opened a new window for the exploration of 
physics beyond the SM~\cite{Giudice:1999ck,Han:1999sg,Uehara:2002yv}, 
in particular, the phenomenology of 
the Higgs sector. In Ref.~\cite{Grzadkowski:1999xh} it is shown that it is possible to 
obtain electroweak symmetry breaking in an extra dimensional scenario even in 
the absence of tree-level Higgs self interactions. Also, in 
Ref.~\cite{edhixcoll} we find scenarios in which the radion in the 
Randall-Sundrum model is contrasted with the SM Higgs boson.
Studying several 
models that lead to a universal rate suppression of Higgs boson observables, 
Ref.~\cite{Wells:2002gq} concluded that the Tevatron and LHC will have 
difficulty finding evidence for extra dimensional effects. Yet 
another study of universal extra dimensions~\cite{Appelquist:2001jz} 
conjectures that a suppression of the Higgs rates occurs~\cite{petriello}.

However, just as in the SM and other four dimensional theories, the Higgs sector 
remains the least constrained, since it can live either on the brane or in 
the bulk, each choice being phenomenologically consistent. One way to 
parametrize this freedom is to consider an extra dimensional two Higgs doublet model 
(XD THDM) where one doublet lives in the bulk while the other is confined to the 
brane. The lightest Higgs boson state, which will resemble the SM one, will then be a 
linear combination of the neutral components of the two doublets. 
Constraints from electroweak precision data have been applied to such a model, 
and it was found that the compactification scale is larger than a couple of 
TeVs~\cite{Masip:1999mk}. 

In this paper, we study the ability of present and future colliders to find the 
lightest Higgs boson in a XD THDM with a single TeV$^{-1}$ size new dimension. 
In particular, our aim is to estimate the minimal size of the compact dimension 
for which the lightest Higgs signal is distinguishable from that of the SM (or 
THDM) at the LHC or at CLIC.

We assume that the SM gauge bosons 
and one of the Higgs doublets propagate in this compact dimension. 
The SM $W^{\pm}$ and $Z$ particles are identified with the zero modes of the 
five dimensional gauge boson fields. There
is a second Higgs field restricted to the brane together with all
the matter fields of the SM. Although the Higgs spectrum includes two
CP-even states ($h,H$, with $m_h < m_H$), one CP-odd Higgs ($A$)
and a charged pair ($H^\pm$), in this work we focus on the lightest
Higgs boson $h$, because most likely this will be the first Higgs state
that future colliders will detect. 
The CP-even Higgses may be the combinations, i.e. 
brane-bulk mixed states, of the two Higgs doublets.

We derive the Lagrangian for Higgs interactions and apply it to
calculate the cross section of the associated production of
Higgs with gauge bosons at the LC, as well as the dominant Higgs decays including the
possible contributions from virtual KK states. 
Crucial to our approach is the cumulative effect of the virtual gauge boson KK 
states, $W^{\pm(n)}$ and $Z^{(n)}$, which contribute to the cross
section for the production of the Higgs associated with the $W^\pm$ and $Z$
and to the three-body decay $h\to Vf\bar{f}'$ ($V=W^{\pm(0)},Z^{(0)}$). 
The corresponding reactions at hadron colliders are studied as well. 

We remark that in this scenario, with a low enough compactification scale, the 
discovery of the extra dimension would probably precede the discovery of the
lightest Higgs boson. Gauge bosons propagating in the new dimension would exhibit 
unambiguous resonances, for example, in their $s$-channel production at the LHC. 
Our focus is on the lightest Higgs because we investigate how much information 
the various colliders can give us about the Higgs sector with a second Higgs 
doublet in the bulk. While more exotic processes might provide more useful to 
this end, we restrict ourselves to $V h$ production ($V=W^\pm$ or $Z$) because 
our emphasis is that this dominant search channel is suppressed at the Tevatron 
(and at LEP), while possibly enhanced at future colliders.

The organization of our paper is as follows. In Section~\ref{sec:modeling}, we
present the model that we use to study the brane-bulk mixing of the
Higgs boson. Then in Section~\ref{production}, we derive formulae for
the Higgs production and decays. These include the evaluation of the
contribution from virtual $Z^{(n)}$ KK states to the associated production at 
linear colliders, i.e. $e^+ e^- \to hZ^{(0)}$, as well as 
to the three-body decay $h\to Vf\bar{f}'$.
Results of our calculation for the hadron collider case are also included. The discussion of the implications of our 
results for present and future colliders appears in 
Section~\ref{sec:Implications}, while the conclusions are presented in 
Section~\ref{sec:conclusion}.

\section{Modeling the Higgs location} 
\label{sec:modeling}

To describe the brane-bulk Higgs mixing, we work with a five dimensional 
(5D) extension of the SM that contains two Higgs doublets.
The SM fermions and one Higgs doublet ($\Phi_u$) live on a 4D boundary,
the brane, while the gauge bosons and the second Higgs doublet ($\Phi_d$),
are all allowed to propagate in the bulk. The constraints from electroweak
precision data~\cite{Masip:1999mk} show that the compactification scale
can be of ${\cal O}$(TeV) (3-4 TeV at 95 \% C.L.).
The relevant terms of the 5D $SU(2)\times U(1)$ gauge and Higgs Lagrangian are 
given by
\be \label{5dlagrangian}
{\cal L}^5 = -\frac{1}{4}\left(F^a_{MN}\right)^2
 -\frac{1}{4}\left(B_{MN}\right)^2
+ |D_M \Phi_d|^2 + |D_{\mu}\Phi_u|^2 \delta(x^5) \, ,
\ee
where the Lorentz indices $M$ and $N$ run from $0$ to $4$, and $\mu$ runs 
from $0$ to $3$. The covariant derivative is given by
\be \label{covariant}
D_M = \partial_M + 
i g_5^{\prime}\frac{Y}{2} B_M + 
i g_5 \frac{\sigma^a}{2} A^a_M \, .
\ee
Given this definition, the mass dimensions of the fields are: 
$\dim(\Phi_d) = 3/2$, $\dim(\Phi_u) = 1$,
$\dim(A_M) = 3/2$, $\dim(B_M) = 3/2$, and the 5D gauge
couplings have a mass dimension of $-1/2$.
Bulk fields are defined to have even parity under $x^5\to -x^5$, 
and are expanded as
\be \label{decomposition}
{\cal S}(x^{\mu},x^5) = \frac{1}{\sqrt{\pi R}}\left(
S^{(0)}(x^{\mu}) + 
\sqrt{2}\sum_{n=1}^{\infty} \cos\left(\frac{nx^5}{R}\right) S^{(n)}(x^{\mu}) 
\right)\, .
\ee
This decomposition, together with Eq.~(\ref{5dlagrangian}), 
guarantees that after compactification we obtain the usual 4D kinetic
terms for all fields. 

Spontaneous symmetry breaking (SSB) of EW 
symmetry occurs when the Higgs doublets acquire vacuum expectation values (vevs). 
After SSB the Higgs fields on the brane can be written as
\bea \label{higgs}
\Phi_u & = & \frac{1}{\sqrt{2}}\left( \begin{array}{c} 
\Phi_u^{0*} \\ \Phi_u^{-} \end{array} \right) =
\frac{1}{\sqrt{2}}\left( \begin{array}{c} 
v_u + h\cos\alpha + H\sin\alpha + i\cos\beta A \\ 
 \cos\beta H^-  \end{array} \right)\, ,
\\
\Phi_d^{(0)} & = & \frac{1}{\sqrt{2}}\left( \begin{array}{c} 
\Phi_d^{+} \\ \Phi_d^{0} \end{array} \right) =
\frac{1}{\sqrt{2}}\left( \begin{array}{c} \sin\beta H^+ \\
v_d - h\sin\alpha + H\cos\alpha + i\sin\beta A \end{array} \right) \, ,
\eea
where the neutral CP-even bosons are denoted by $h$ and $H$, and $h$ is identified 
with the lightest Higgs: $m_h < m_H$.
The mixing angle $\alpha$ is introduced to diagonalize the CP-even mass matrix.
The CP-odd and charged Higgs fields are denoted by $A$ and $H^\pm$, and $v_u$ 
and $v_d$ are the vevs of $\Phi_u$ and $\Phi_d^{(0)}$ respectively. Note that 
the angles $\alpha$ and $\beta = \arctan(v_u/v_d)$ parametrize what we call 
brane-bulk mixing, or higher dimensional `location', of the neutral Higgses.

After performing the KK-mode expansion and identifying the
physical states, one derives the interaction Lagrangian for all 
the vertices of the neutral and charged Higgses. In particular,
the interactions $ZZh$ and $ZZ^{(n)}h$, which are necessary to 
calculate the Higgs production in association with a $Z$, 
as well as the vertices involving the $W^{\pm}$
bosons, are given by the following 4D Lagrangian:
\bea \label{interactions}
\nonumber
{\cal L}^4 & \supset & \frac{g M_Z}{2c_W}\left(h\sin(\beta-\alpha)
+H\cos(\beta-\alpha)\right)Z_{\mu}Z^{\mu} \\ \nonumber
           &    +    &
\sqrt{2} \frac{g M_Z}{c_W} \left( h \sin\beta\cos\alpha
+H\sin\beta\sin\alpha \right)\sum_{n=1}^{\infty} Z_{\mu}^{(n)}Z^{\mu}
\\ \nonumber
           &    +    & g M_W \left(
h\sin(\beta -\alpha) +H\cos(\beta -\alpha)\right)
W^+_{\mu}W^{-\mu} \\  
           &    +    &
\sqrt{2}gM_W\left(h\sin\beta\cos\alpha + H\sin\beta\sin\alpha\right)
\sum_{n=1}^{\infty}\left(W^+_{\mu}W^{-(n)\mu} + 
W^-_{\mu}W^{+(n)\mu}\right) \, .
\eea
%where $\tan\beta = v_u/v_d$.
Thus, the vertices $hZZ$ and $hWW$ have the same form as in the usual 4D THDM, 
i.e. proportional to $\sin(\beta-\alpha)$.
Meanwhile the couplings $hZZ^{(n)}$ and $hW^{\pm}W^{\pm(n)}$ are proportional to 
$\sin\beta\cos\alpha$, vanishing either when $\beta=0$ or $\alpha = \pi/2$, 
i.e. either when EWSB is driven exclusively by the vev of $\Phi_d^{(0)}$ or
when the CP-even Higgs comes entirely from $\Phi_d^{(0)}$. 
Similarly, the couplings of the CP-odd Higgs $A$ and the charged Higgs 
resemble the THDM, although new vertices of the type $H^+W^-Z^{(n)}$ or 
$H^+W^{-(n)}Z$ could be induced.

On the other hand, because the fermions are confined to the brane,
the Higgs-fermion couplings could take any of the
THDM I, II or III versions~\cite{Diaz-Cruz:wp,Diaz-Cruz:1999xe}. 
However, for the THDM III version 
the possible FCNC problems would be ameliorated, as the bulk-brane couplings 
will be suppressed by 
the factor $1 / \sqrt{2\pi R}$~\cite{Sakamura:1999pt}. 
Thus, for the flavor conserving
couplings one can use the formulae of the widely studied THDM II, which
appears for instance in the Higgs Hunters Guide~\cite{Gunion:1989we}.

\section{Extra dimensional contribution to Higgs production and decays} 
\label{production}

\subsection{Associated production $h+Z$ at linear colliders}

In order to study Higgs production at future colliders, first 
we derive the cross section for the Bjorken process, namely for
$e^+e^-\to hZ$. The total amplitude includes the contribution of
virtual $Z=Z^{(0)}$ and $Z^{(n)}$ states in the $s$-channel. 
The sum over all KK modes can be performed analytically, which considerably
simplifies the final expression. Our result for the cross section is given by
\bea \label{cross} 
\sigma(e^+e^- \rightarrow hZ) = \sigma_{SM} F_{XD}(\alpha,\beta,s) \, .
\eea
Here $\sigma_{SM}$ denotes the SM cross 
section, given by
\be \label{crossSM}
\sigma_{SM} = \frac{G_F^2 M_Z^4}{3\pi} \left(4s_w^4-2s_w^2+\frac{1}{2} \right)
\frac{|{\bf k}|}{\sqrt{s}}\frac{(3M_Z^2+|{\bf k}|^2)}{(s-M_Z^2)} \, ,
\ee
with
\bea 
|{\bf k}| = 
\frac{1}{\sqrt{s}}\left(\left(\frac{s+M_Z^2-m_h^2}{2}\right)^2 
-s M_Z^2\right)^{1/2}
\eea
being the 3-momentum of the $Z$ boson.

The extra dimensional contribution is factorized into
\bea \label{FXD}
F_{XD}(\alpha,\beta,s) = 
\left[ \sin(\beta-\alpha) + 2\cos\alpha\sin\beta F_{KK}(s) \right]^2  \, .
\eea
The function $F_{KK}$, which arises after summing over all the virtual KK-modes, 
is given by
\be \label{fkk}
F_{KK}(s) = 2 \sum_{n=1}^{\infty} \frac{s-M_Z^2}{s-M_n^2} 
          = R A(s) \pi \cot(R A(s) \pi) - 1 \, ,
\ee
where 
\be \label{Mn}
M_n = \sqrt{n^2/R^2 + M_Z^2} \, ,
\ee
is the mass of the $n^{th}$ KK level. When neglecting the widths, 
\be \label{width}
\Gamma_n = M_n \alpha_g(v_f^2+a_f^2)/3 \, ,
\ee
of the KK resonances\footnote{The width of the $n$-th KK state is defined as 
its total decay rate into a SM fermion pair.}, 
we can write 
\be \label{As}
A(s) = \sqrt{s-M_Z^2} \, . 
\ee
(Here $v_f$ and $a_f$ are the SM vector and axial coupling strength of the 
vector boson to fermions.) This is a reasonable approximation, since $\alpha_g = 
g^2/(4 \pi) \ll 1$. By neglecting only $M_Z \alpha_g$ terms (next to $M_Z$), we 
can even include the dominant width effect by setting 
\be \label{Asc}
A(s) = \sqrt{c(s-cM_Z^2)}/c \, , 
\ee
where $c = (1+2i\alpha_g)$. 

\begin{figure}
\includegraphics[width=12cm]{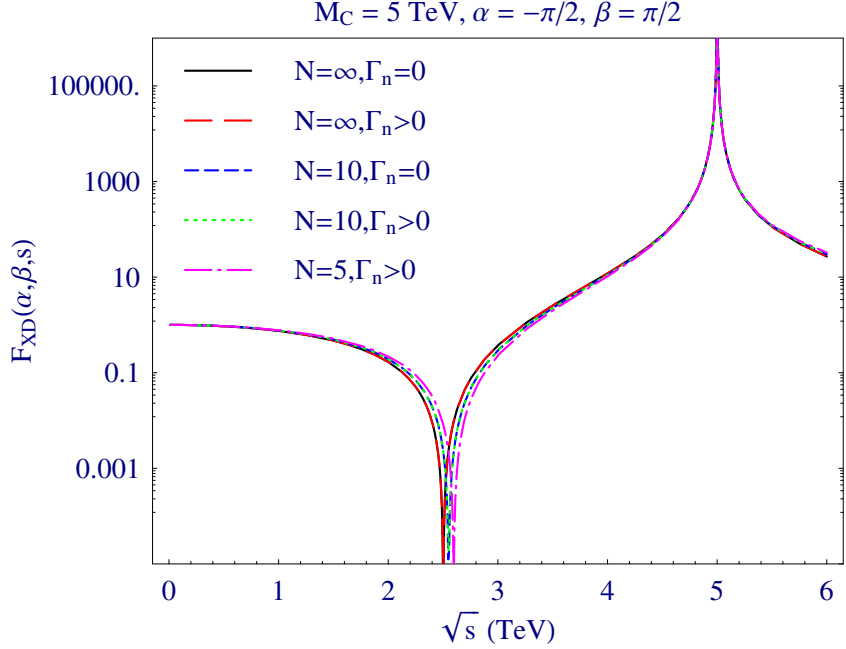}
\caption{The extra dimensional contribution $F_{XD}$ to $e^+e^- \to Z h$, as 
the function of the center of mass energy $\sqrt{s}$ for various other 
parameter values.}
\label{fig:FXD}
\end{figure}

Recently it was pointed out that summing over a large number 
of KK resonances may jeopardize the unitarity of standard-like extra dimensional 
models \cite{Chivukula:2003kq}. Thus, we mention that the sum in Eq.(\ref{fkk}) 
can also be performed analytically for a finite number of terms with the 
result:
\be \label{fkkN}
F_{KK}(s) = 2 \sum_{n=1}^{N} \frac{s-M_Z^2}{s-M_n^2} 
          = R A(s) (\pi \cot(R A(s) \pi) - H_{N-RA(s)} + H_{N+RA(s)}) - 1 \, ,
\ee
where $H_{x}$ is the harmonic number function. (The KK width can also be 
included as above). Since the $F_{KK}$ function can be calculated even 
analytically, we can easily check its sensitivity to the number of KK levels and 
the inclusion of KK width. Fig.(\ref{fig:FXD}) shows that, for a typical set of 
parameters, $F_{KK}$ is reasonably insensitive to these, which ensures that our 
later results are robust against cutoff and width effects of KK levels in the 
relevant energy range.

These expressions also apply for the process $q{\bar q}' \to W^{\pm}h$ after 
changing $\sigma_{SM}$ and $M_Z$ to $M_W$ at the appropriate places.

\subsection{Associated production $h+Z, \, h+W^\pm$ at hadron colliders}

When considering Higgs production at hadron colliders, an expression similar to 
Eq.(\ref{cross}) holds at the parton level for the production cross section of 
the Higgs in association with a $W^\pm$ or $Z$. To obtain the hadronic cross 
section $h_1 h_2 \to hZ$, the partonic cross section must be convoluted with 
the parton distribution functions (PDFs):
\be
\sigma(h_1 h_2 \rightarrow Z h) = 
\sum_{q \bar{q}} \int_{0}^{1} \int_{0}^{1}
f_{q/h_1}(x_1,\hat{s}) 
\sigma(q \bar{q} \rightarrow Z h) %(x_1,x_2,\hat{s})
f_{\bar{q}/h_2} (x_2,\hat{s}) d x_1 d x_2
+ {q \leftrightarrow \bar{q}} \, .
\label{hadrXS}
\ee
Here $f_{q/h_i}(x,{\hat s})$ gives the distribution of a parton $q$ in the 
hadron $h_i$ as a function of the longitudinal momentum fraction $x$ and the 
factorization scale, which is chosen to be the partonic center of mass 
$\hat{s}$. In our numeric study we use CTEQ4M PDFs~\cite{Lai:1996mg}. The sum in 
Eq.~(\ref{hadrXS}) extends over the light quark flavors $q = u,d,s,c$.

The large center of mass energy that can be achieved at the LHC also opens up 
the possibility to produce a Higgs boson in association with KK states, for 
instance $hZ^{(1)}$, which will have a very distinctive signature that could 
allow `direct' detection of the first KK modes at the LHC. This possibility is 
studied elsewhere.

\subsection{Higgs decays}

For Higgs bosons lying in the intermediate
mass range, which is in fact favored by the analysis of electroweak 
radiative corrections, the dominant decay is into $b\bar{b}$ pairs.
In our higher dimensional model this decay width is given by the formulae of the
THDM, just as that of the other tree-level two-body modes. On the other hand, for
the three-body decays $h \to W l \nu_l$ and $h\to Z l^+l^-$, which can play 
a relevant role at the Tevatron and LHC, the corresponding decay width could
receive additional contributions from the virtual $KK$ states. 
The inclusion of these KK modes leads to the following
expression for the differential decay width:
\be \label{gamma}
\frac{d\Gamma}{dx} (h\to W l\bar{\nu}_l)= \frac{g^4m_h}{3072 \pi^3}
\frac{(x^2-4r_w)^{1/2}}{1-x} f_V(x)
\left[ \sin(\beta-\alpha) + 2 \cos\alpha\sin\beta F_{KK} \right]^2 \, ,
\ee 
where
$f_V(x)= x^2-12r_w x +8r_w +12r^2_w$, with $r_w=M_W^2/m_h^2$,
and $2r^{1/2}_w < x <1+r_w$. The $F_{KK}$ function
is given as in Eq.~(\ref{fkk}), with the replacements 
$s\to q^2=m_h^2 (1-x)$ and $M_Z \to M_W$. 
A similar expression can be derived for the
decay $h\to  Z l^+l^-$.

To study the effect of the KK modes on the decay $h\to W l\bar{\nu}_l$, 
we have evaluated the ratio of the corresponding decay width in the
extra dimensional scenario over the SM decay width:
\be
R_{hWW^*}=\frac{\Gamma( h\to W l\bar{\nu}_l)_{XD}}
               {\Gamma( h\to W l\bar{\nu}_l)_{SM}} \, .
\label{RhWW}
\ee
Results for this ratio are shown in Table~\ref{tab:decays}, for 
several representative sets of parameters which are chosen as
\begin{itemize}
 \item[{\bf A}]: $M_c = 2$ TeV, $\alpha = \pi/3$, ~~~ 
       {\bf B} : $M_c = 2$ TeV, $\alpha = \pi/1.28$,
 \item[{\bf C}]: $M_c = 2$ TeV, $\alpha = \pi$,   ~~~~~~
       {\bf D} : $M_c = 5$ TeV, $\alpha = \pi$, 
\end{itemize}
while $\beta=\pi/4$ remains fixed. 
We observe that significant deviations from the SM can appear, although this
effect is largely due to the difference of the Higgs couplings in the THDM and
the SM.
%One can appreciate that significant deviations
%from the SM can appear, though the larger effect comes from the 
%deviations for the THDM Higgs couplings from the SM case.

\begin{table}
\begin{tabular}{|c|c|c|c|c|}
\hline
~$m_h$ (GeV)~ & ~$R_{hWW^*}$ (set {\bf A})~ & ~$R_{hWW^*}$ (set {\bf B})~ & 
~$R_{hWW^*}$ (set {\bf C}) ~& $R_{hWW^*}$ (set {\bf D})~ \\ 
\hline
 130   & $6.0\times 10^{-2}$ & 0.999 & 0.51 & 0.50 \\
\hline
 140   & $6.0\times 10^{-2}$ & 0.998 & 0.51 & 0.50 \\
\hline
 150   & $6.1\times 10^{-2}$ & 0.996 & 0.50 & 0.50 \\
\hline
 160   & $6.1\times 10^{-2}$ & 0.993 & 0.50 & 0.50 \\
\hline
\end{tabular}
\caption{The ratio $R_{hWW^*}$, introduced in Eq.~(\ref{RhWW}), for several sets 
of parameters {\bf A}, {\bf B}, {\bf C}, {\bf D} (as defined in the text), and 
with $\beta=\pi/4$.}
\label{tab:decays}
\end{table}

On the other hand, the loop induced decays $h\to \gamma \gamma, Z \gamma$
also receive contributions from the $W^{(n)}$ KK modes. But since the
coupling $hWW^{(n)}$ that appears in the loop is proportional to $M_W$, rather
than $M_{W^{(n)}}$, the contribution of the KK states will decouple, as there 
are no mass factors that could cancel the ones in the numerator. 
Thus, the KK contribution can be neglected for the 
decay widths of the loop induced decays.

We conclude that in the intermediate Higgs mass range
%, which is in fact the one preferred by the analysis of EW precision measurements, 
the decay $h\to b\bar{b}$ will continue to dominate, even more than in the SM case
for some values of parameters. In the limit $M_c >> O(1)$ TeV, one recovers
the SM pattern for the Higgs decays.

\section{Implications for Higgs searches at future colliders} 
\label{sec:Implications}

\subsection{The LC case}

\begin{figure}
\vspace{-1cm}
%  \begin{centering}
%  \def\epsfsize#1#2{1.0#2}
\resizebox*{.45\textwidth}{.35\textheight}
%\hfil\hspace{-10em} 
{\includegraphics{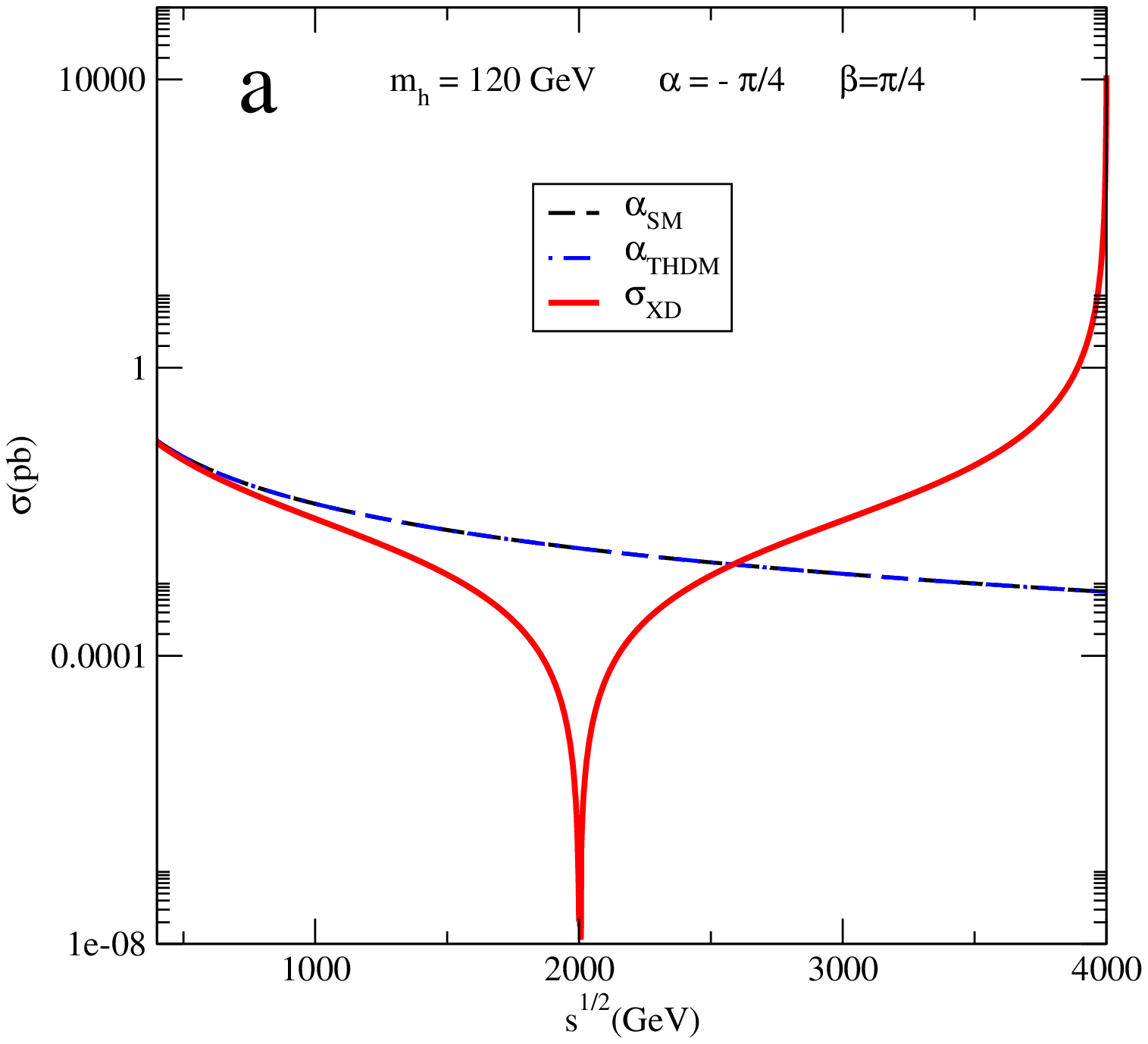}}
\resizebox*{.45\textwidth}{.35\textheight} 
{\includegraphics{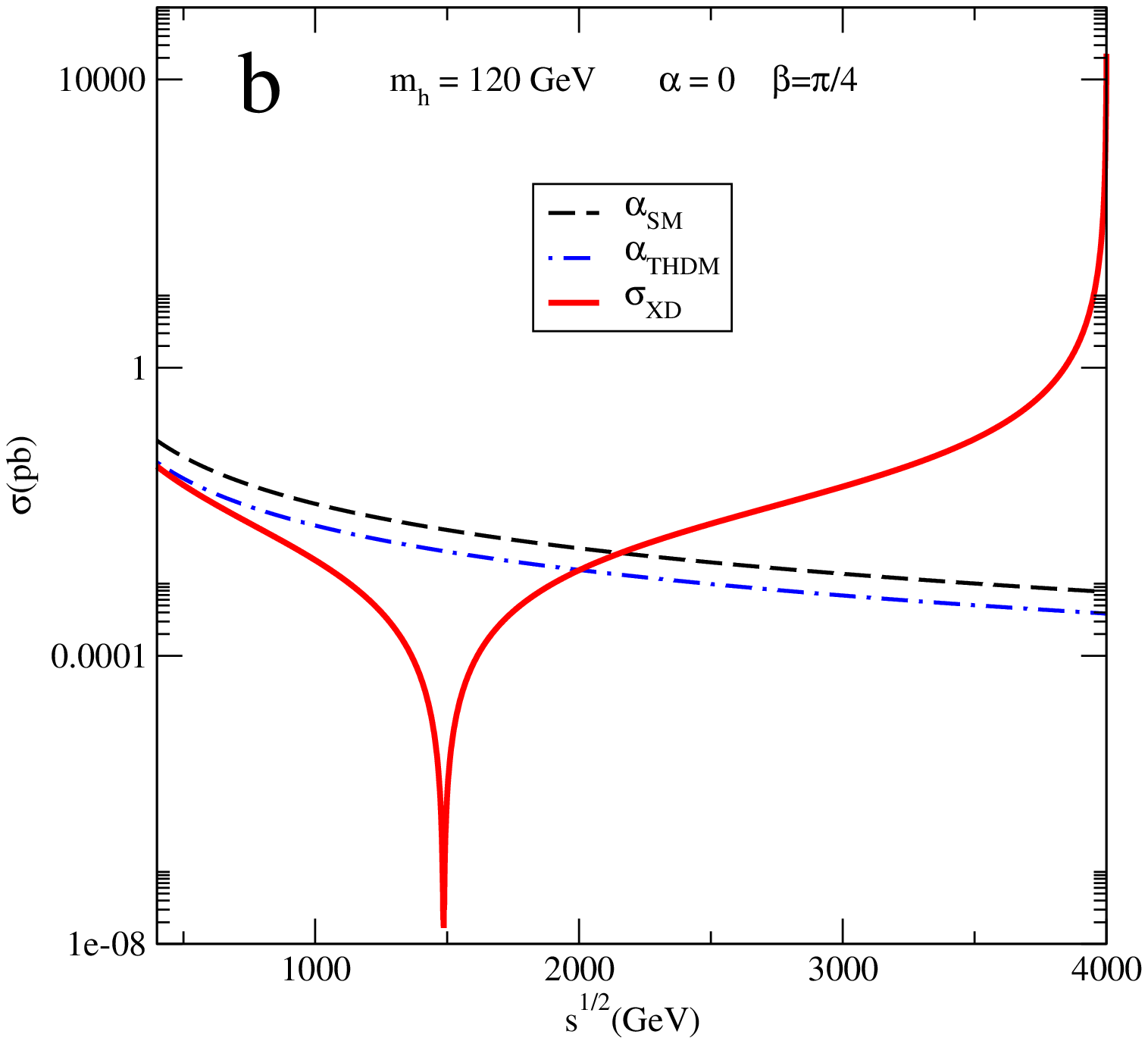}}\\
\resizebox*{.45\textwidth}{.35\textheight} 
{\includegraphics{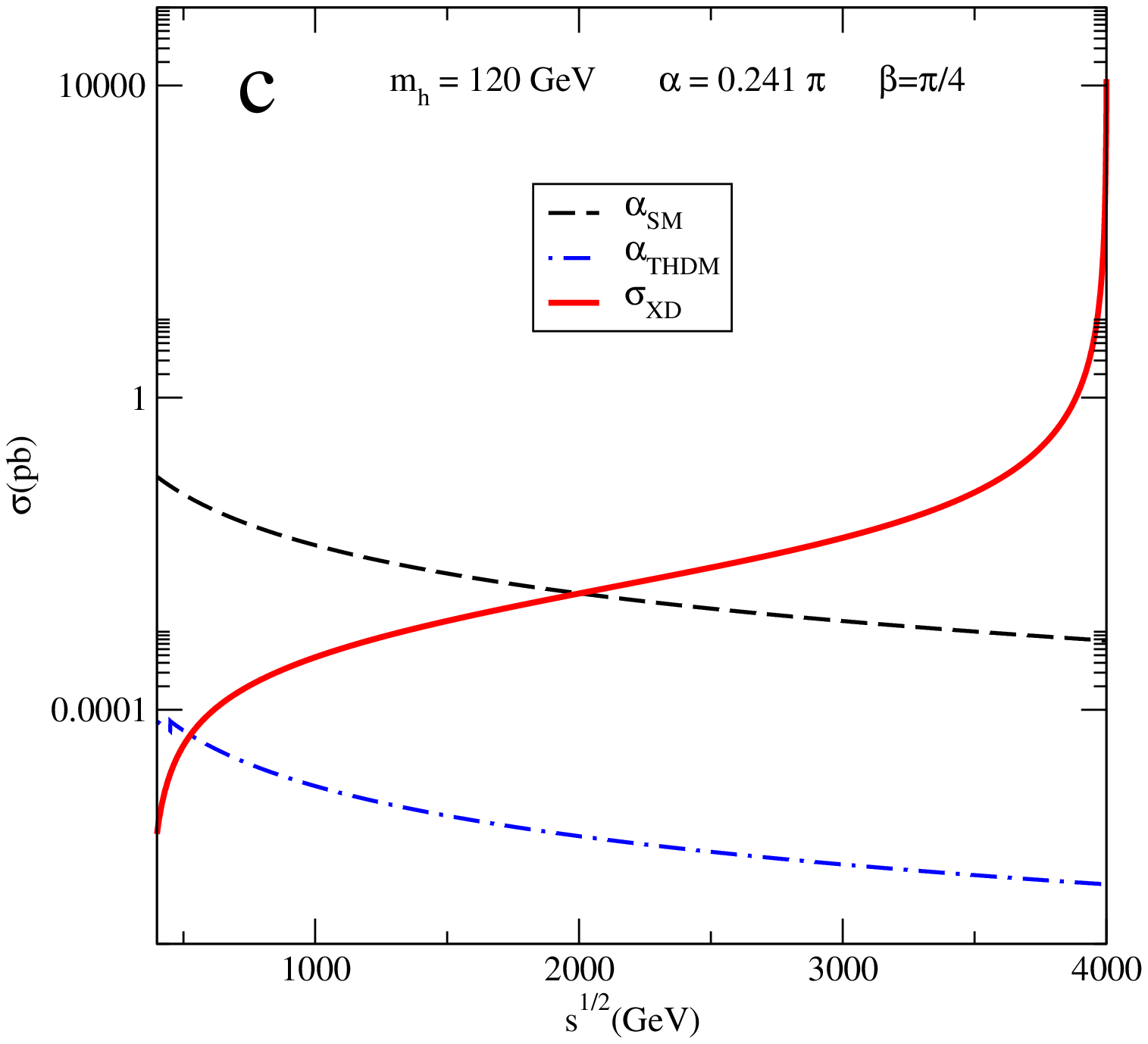}}
%\epsfbox{fig1a.ps} \hfill
\caption{SM, THDM and XD cross sections for $e^+e^- \rightarrow hZ$. Each plot
corresponds to a different set of values for $\alpha$ and $\beta$ all with
$m_h = 120$~GeV and with a compactification scale $M_c = 4$~TeV.}
\label{figure1}
%  \end{centering}
\end{figure}

Because of its simplicity, first we discuss Higgs production at a linear 
collider. The present bound on the compactification scale is $M_c \gtsim 
3.8$~TeV~\cite{Rizzo:1999br}, for the cases when the Higgs field is either in 
the bulk or confined to the brane. For the general case of ``mixing'' one 
obtains similar bounds~\cite{Muck:2002af}.

The results for the $e^+e^- \to hZ$ cross section, after the inclusion 
of the virtual $Z$ KK contribution, are shown in Fig.~\ref{figure1}.
This plot shows the cross section as a function of the center of mass 
energy ($200 < \sqrt{s} < 4000$ GeV),  for $m_h=120$ GeV and a value of the 
compactification scale $M_c=4$ TeV. We plot the SM cross section $\sigma_{SM}$, the
THDM cross section $\sigma_{THDM}$, as well as the extra dimensional cross section
$\sigma_{XD}$. Three different pairs of $\alpha$ and $\beta$ were chosen to 
compute $\sigma_{THDM}$ and $\sigma_{XD}$. They are given by
\begin{enumerate}
 \item[{\bf a}]: $\beta=\pi/4$, \, $\alpha=-\pi/4$,  ~~~\,
       {\bf b} :  $\beta=\pi/4$, \, $\alpha=0$, ~~~\,
       {\bf c} : $\beta=\pi/4$, \, $\alpha=0.241\pi$.
%       {\bf d} :  $\beta=\pi/3$, \, $\alpha=0.32\pi$. 
\end{enumerate}
Choice {\bf a} corresponds to a case in which $\sigma_{THDM} = \sigma_{SM}$; 
{\bf b} corresponds
to a case where the effect of the mixing due to $\alpha$ in the term involving the
KK sum is maximum; {\bf c} corresponds to values of $\alpha$ and $\beta$ 
for which $\sigma_{XD}/\sigma_{THDM}$ is larger than 1 above $\sqrt{s}=2$~TeV. 

The first frame of Fig.(\ref{figure1}) shows well that the shape of the cross
section is determined by the product of $F_{KK}$, as shown in 
Fig.(\ref{fig:FXD}), and the SM cross section.

We can see that in all three cases the cross section of the XD model is always
smaller than that of the SM at $\sqrt{s} = 500, 1000$~GeV, and that in order to 
obtain a larger cross section, one needs energies greater than 
$\sqrt{s} \sim 2$~TeV. This is understood from the fact that the
heavier KK modes, through their propagators, interfere destructively with the SM 
amplitude thus reducing the cross section. 
%It is only until one reaches energies close to the
%first KK mass that an enhancement is observed due to the dominating KK 
%contribution. 
We mention that the three cases presented in Fig.~\ref{figure1} are only 
representative, 
and that one can find broad regions of parameter space in which 
$\sigma_{XD} > \sigma_{THDM}$.

Moreover, as Fig.~\ref{figure1} shows, once the center of mass energy 
approaches the threshold for the production of the first KK state, 
the cross section starts growing. For instance, with $M_c=4$ TeV,  
$\sigma_{SM} \simeq \sigma_{XD}$ for $\sqrt{s} \simeq 2$ TeV.
However, one would need higher energies in order to have a cross 
section larger than that of the SM, which may only be possible at 
CLIC~\cite{Assmann:2000hg}.

According to current studies, when the cross section is 4\% larger than the SM 
cross section, with the estimated precision that could be obtained at the 
LC~\cite{Battaglia:2000jb}, it may be possible to distinguish between the SM and 
XD Higgs scenarios. We can see that this might be possible at CLIC for 
$M_c = 3$--4 TeV for a broad range of parameter values.
It is interesting to note that such deviations in the cross section from the SM 
prediction arise even when the couplings of the Higgs to the gauge bosons are 
indistinguishable from the SM couplings.

\subsection{Implications for the Tevatron}

After the productive but unsuccessful Higgs search at LEP2, the Run 2 of the 
Tevatron continues the search until the LHC starts operating. The luminosity 
that is required to achieve a 5 or 3 $\sigma$ discovery, or a 95\% C.L. 
exclusion limit, was presented by the Run 2 Higgs working group 
\cite{Carena:2000yx}. For instance, with $m_h=120$ GeV, the corresponding 
numbers are about 20, 6 and 2 $fb^{-1}$ respectively.

Assuming $M_c \ge 2$ TeV, the inclusion of the KK modes decreases the 
$h W^\pm$ and $h Z$ associated production cross section at the Tevatron.
Depending on the actual values of $M_c$, $\alpha$ and $\beta$, the suppression in 
the parton level cross section may be anything between 1 and 99 percent. For 
example, if $M_c$ is a few TeV then the $s$-channel process $q{\bar q'} \to 
W^{\pm}h$ can receive a considerable suppression, as it can be inferred from 
Fig.~\ref{figure1}. This is so unless the KK contribution 
itself is suppressed by $\cos\alpha$ and/or $\sin\beta$ in Eq.~(\ref{cross}), 
in which case the presented model has little relevance.
Thus, as a general prediction of this model, we conclude that more than the 
above listed luminosity is required to find a light Higgs boson. This
slims the chances of the Tevatron to find the Higgs of this model.

% We note that, the Higgs boson can also be produced via gluon fusion, and for 
% values of the Higgs mass where the decay mode $h\to Wl\nu_l$ can be used, we 
% need to include also the corrections to the decay width  $\Gamma (h \to 
% WW^*)=Z_2 \Gamma_{SM} $. It turns out that the resulting number of events will 
% scale like $N_{SM} \to Z_1 Z_2 N_{SM}$. Thus, the required luminosity to get a 
% fixed number of events will scale like ${\cal{L}} \to {\cal{L}}/(Z_1Z_2)$, which 
% can produce interesting effects...

\subsection{Higgs production at the LHC}

\begin{figure}
\includegraphics[width=12cm]{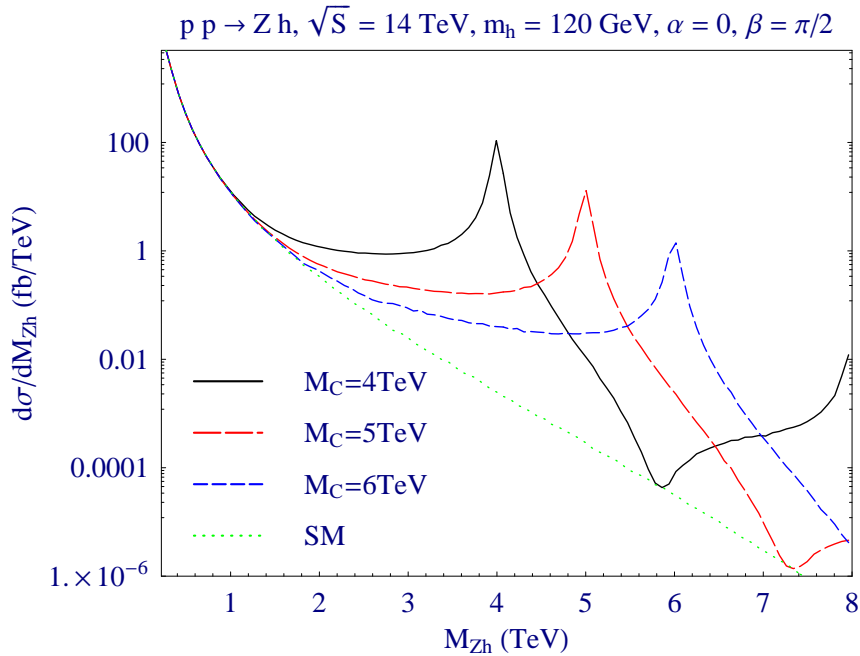}
\caption{Higgs production cross section in association with a $Z$ boson at the
LHC as a function of the compactification scale for selected values of the 
mixing parameters.}
\label{figLHC1}
\end{figure}

The Higgs discovery potential in this model is more promising at the LHC. We 
illustrate this in Fig.~\ref{figLHC1}, showing the $pp\to hZ$ differential cross 
section as the function of the $hZ$ invariant mass $M_{hZ}$. 
The typical resonance structure displayed by Figs.~\ref{fig:FXD} and 
\ref{figure1} is preserved by the hadronic cross section. The resonance peak
is well pronounced when $M_{hZ} \sim M_c$. This leads to a large enhancement 
over the SM (or THDM) cross section.

% enhancement
The singularity at $M_c = M_{hZ}$ is regulated by the width of the KK mode, 
which is included in our calculation as given by Eq.(\ref{width}). Thus, 
Fig.~\ref{figLHC1} gives a reliable prediction of the XD cross section even in 
the peak regions. Depending on the particular values of $\alpha$ and $\beta$ the 
enhancement is more or less pronounced. For an optimistic set $\alpha = 0$ and 
$\beta = \pi/2$, the XD production cross section is considerably enhanced 
compared to the SM at $M_{hZ} = M_c$. This enhancement may be detectable up to 
about $M_c = 6$ TeV. We estimate that with 100 fb$^{-1}$ for $M_c = 6$ TeV there
are about 20 $hZ$ events in the bins around $M_{hZ} \sim 
M_c$. As Fig.~\ref{figLHC1} shows, in the SM less than one event is expected in 
the same $M_{hZ}$ range. It is needless to say that similar results hold for 
$pp\to hW^\pm$, which further enhances the discovery prospects.

% An alternative way to gauge the XD enhancement is shown in Fig.~\ref{figLHC2}. 

% detection 
Based on these results, we conclude that in the Bjorken process alone the reach 
of the LHC may extend to about $M_c = 6$ TeV, depending on the values of 
$\alpha$ and $\beta$.
Finally, we note that the XD contribution to the running of the gauge couplings 
is important when the effective center of mass energy of the collider is close 
to $1/R$~\cite{Dienes:1998vh,Dienes:1999vg}. Since not included in this work, 
this contribution is expected to change our results somewhat for 
$\sqrt{\hat s} \gtsim M_c$ .

\section{Conclusions} 
\label{sec:conclusion}

In this work, we studied the extent to which present and future colliders can 
probe the brane-bulk location of the Higgs boson in a model with a TeV$^{-1}$ 
size extra dimension. In this model one Higgs doublet is located on the brane 
while another one propagates in the bulk.
We found that the virtual KK states of the gauge bosons contribute to the 
associated production of the Higgs with $W^{\pm(n)}$ and $Z^{(n)}$, and at low 
energies ($\sqrt{s} < 1/R$~GeV) the cross section is suppressed compared to 
the SM case. Meanwhile at higher energies, i.e. at $\sqrt{s} \sim 1/R$, the cross 
section can receive an enhancement that has important effects on the discovery 
of the Higgs at future colliders. 

Analysing compactification scales in the range of 2--8 TeV, we concluded that to 
find a Higgs signal in this model the Tevatron Run 2 and the LC with 
$\sqrt{s}=500-1500$ GeV are required to have a luminosity higher than in the SM 
case. 
Meanwhile, the LHC and possibly CLIC with $\sqrt{s} \sim 3$--5 TeV might have a
greater potential to find and study a Higgs signal. Depending on the model 
parameters, these colliders may be able to distinguish between the extra 
dimensional and the SM for compactification scales up to about 6 TeV. If this 
model is relevant for weak scale physics, the LHC should see large enhancements 
in the associated production rates. Thus, not finding the Higgs at the Tevatron 
may be good news for the XD Higgs search at the LHC.

\section*{Acknowledgments}

A.A. was supported by the U.~S. Department of Energy under grant 
DE-FG02-91ER40676.
C.B. was also supported by the DOE, under contract number 
DE-FG02-97ER41022.
J.L. D.-C. was supported by CONACYT and SNI (M\'exico).

\end{document}